\documentclass[aps,prd,showpacs,showkeys,superscriptaddress,nofootinbib,twocolumn]{revtex4-1}
\usepackage[colorlinks,linkcolor=blue,anchorcolor=blue,citecolor=blue,urlcolor=blue,breaklinks=true]{hyperref}
\usepackage{graphicx}
\usepackage[tbtags]{amsmath}
\usepackage{amssymb}
\usepackage{slashed}
\usepackage{latexsym}
\usepackage{epsfig}
\usepackage{amsbsy}
\usepackage{array}
\usepackage{amssymb}
\usepackage{setspace}
\usepackage{bm}
\usepackage{lipsum}
\usepackage{mathrsfs}
\usepackage{float}
\usepackage{color}
\usepackage{booktabs}
\usepackage[T1]{fontenc}
\usepackage{mathptmx}
\DeclareMathAlphabet{\mathcal}{OMS}{cmsy}{m}{n}
\DeclareSymbolFont{largesymbols}{OMX}{cmex}{m}{n}

\begin{document}
\author{Li-Kang Yang}
\email{yanglk@smail.nju.edu.cn}
\affiliation{Department of physics, Nanjing University, Nanjing 210093, China}
\author{Xiaofeng Luo}
\email{xfluo@mail.ccnu.edu.cn}
\affiliation{Key Laboratory of Quark $\&$ Lepton Physics (MOE) and Institute of Particle Physics, Central China Normal University, Wuhan 430079, China}
\author{Hong-Shi Zong}
\email{zonghs@nju.edu.cn}
\affiliation{Department of physics, Nanjing University, Nanjing 210093, China}
\affiliation{Nanjing Proton Source Research and Design Center, Nanjing 210093, China}
\affiliation{Joint Center for Particle, Nuclear Physics and Cosmology, Nanjing 210093, China}
\date{\today}

\title{QCD phase diagram in chiral imbalance with self-consistent mean field approximation}

\begin{abstract}
We employ a new self-consistent mean field approximation of NJL model, which introduces a free parameter $\alpha$ ($\alpha$ reflects the weight of different interaction channels), to study the effects of the chiral chemical potential $\mu_5$ on QCD phase structure, especially the location of the QCD critical end point (CEP). We find that, at a high temperature, the critical temperature of QCD phase transition smoothly increases with $\mu_5$ at the beginning, and then decreases rapidly. At low temperature and high baryon density region, the increase of the chiral chemical potential will reduce the critical chemical potential of phase transition. The temperature of the CEP shows a non-monotonic dependence on the chiral chemical potential with a long plateau around the maximum. At $\mu_5=0$, we found that the CEP will disappear when the $\alpha$ value is larger than 0.71 and will reappear when the $\mu_5$ increases. This study is important for exploring the QCD phase structure and the location of CEP in chiral imbalanced systems.

\end{abstract}


\maketitle
\section{Introduction}
As a fundamental theory of strong interaction, Quantum Chromodynamics (QCD) plays an important role in the Standard Model. At the low energy scale, the most prominent features of the QCD vacuum are color confinement and spontaneously chiral symmetry broken. As the temperature and baryon chemical potential increase, the QCD matter will undergo a phase transition from hadronic phase to quark-gluon plasma (QGP), which is a deconfined and approximate chiral symmetric state \cite{Stephanov:2007fk,PhysRevLett.34.1353,MeyerOrtmanns:1996ea,Martinez:2013xka}. The QGP is expected to exist in the early universe and can be created in relativistic heavy-ion collisions.

Because of the non-Abelian feature of the QCD gauge group, the vacuum has many topological structures. Each of the vacuum state can be characterized by an integer number called Chern-Simons charge \cite{Belavin:1975fg,PhysRevLett.37.172,Chern:1974ft}. In the domain of non-perturbative QCD, the instantons and sphalerons with wind number $Q_w\ne 0$ could change the topological structure of the vacuum \cite{PhysRevLett.37.172,tHooft:1976rip,*PhysRevD.18.2199.3,*PhysRevD.14.3432}. Because of the high potential barrier of scale $\Lambda_{QCD}$ between the two different vacuum states, at the low temperature, the transition between them relies on the instantons tunneling \cite{PhysRevD.14.3432}. Therefore the transition rate is largely limited. When the temperature reaches the scale $\Lambda_{QCD}$ of the QGP phase, the sphalerons can leap over the barrier, which will lead to the significant increasing of the transition rate \cite{Kuzmin:1985mm,Shaposhnikov:1987tw,Arnold:1987mh,Arnold:1987zg}. When the quarks interact with these topological gauge fields, the helicities of the quarks will change, which results in the chiral imbalance between left- and right-hand quarks \cite{Adler:1969gk,Bell:1969ts}:
\begin{equation}
N_L-N_R=2N_fQ_w.
\end{equation}
This chiral imbalance would lead to a local $P$ and $CP$ violation \cite{PhysRevLett.81.512,Kharzeev:2007jp,Kharzeev:2009fn}. Because of the higher temperature and the approximation chiral restoration of the QGP phase, the chiral imbalance is more obvious there. In the case of the strong magnetic field, the right- and left-hand quarks move in different directions along the magnetic field, and the chiral imbalance would result in the observable effects in experiment \cite{Burnier:2011bf,Son:2004tq,Fukushima:2008xe}.

Due to the complexity and non-perturbative feature of  QCD, it is a big challenge to map out the QCD phase structure. The first principle Lattice QCD calculation has confirmed that the nature of the QCD phase transition at zero baryon chemical potential is a smooth crossover \cite{Aoki:2006we}.  In the case of finite baryon chemical potential $\mu_B$, the Lattice QCD meets the sign problem.  Thus, there are still large uncertainties in determining the QCD phases structure at high baryon density region from theoretical side.

Experimentally, relativistic heavy-ion collisions can create extreme environment and provide us a useful tool to study the phase structure of hot dense nuclear matter in a controllable way.   By tuning the colliding energies, one can scan the QCD phase diagram and search for the possible signatures of QCD critical point and/or first order phase transition \cite{Luo:2017faz,PhysRevLett.105.022302,*PhysRevLett.112.032302,Luo:2015doi,Adamczyk:2017wsl}. It is the main goal of the Beam Energy Scan program (BES) program (BES-I$\&$II: 2010-2021) conducted at Relativistic Heavy-Ion Collider (RHIC). At the beginning of the non-central heavy-ion collisions, a strong magnetic field of $10^{14}$ Tesla will be generated \cite{Kharzeev:2007jp,Kharzeev:2004ey}. Considering the chiral imbalance of the QGP, it will result in an observable electric current along the magnetic field direction. This phenomenon is called chiral magnetic effect (CME) \cite{Fukushima:2008xe,Kharzeev:2013ffa,Kharzeev:2015znc}. The observation of CME can be seen as evidence for the local $CP$ violation and the existence of non-trivial gauge configurations. Therefore, in order to better understand the relativistic heavy-ion collision experiment, it is important for us to study the influence of the chiral imbalance on the strong interaction phase diagram.

In order to investigate the system where right- and left-hand quark are asymmetric, we can introduce the chiral chemical potential $\mu_5$  \cite{Ruggieri:2011xc,Fukushima:2008xe} which connects to the chiral number density $n_5=n_R-n_L$. Just as the chemical potential $\mu$ can reflect the density of the quark, we introduce the chiral chemical potential to denote the imbalance between the right- and left-hand quarks. Since the real state of QGP is chiral imbalanced, we need to consider the influence of $\mu_5$ when we study the phase transition of the strong interaction. In previous works, people have studied the effects of $\mu_5$ on the phase transition and critical end point (CEP) with the PNJL model \cite{Fukushima:2010fe,Cui:2016zqp,Pan:2016ecs,Ruggieri:2011xc}, Dyson-Schwinger equation \cite{Wang:2015tia,Cui:2016zqp,Xu:2015vna,Tian:2015rla} and quark-meson model \cite{Ruggieri:2011xc} among others.

The standard Lagrangian of NJL model contains scalar $(\bar{\psi}\psi)^2$ and pseudoscalar-isovector $(\bar{\psi} i \gamma_5\boldsymbol\tau \psi)^2$ channels. We can produce not only scalar $(\bar{\psi}\psi)^2$ and pseudoscalar-isovector $(\bar{\psi} i \gamma_5\boldsymbol\tau \psi)^2$ through standard Fierz transformations, but also generate other interaction channels. These interaction channels will play an important role in the case of specific background fields. For example, in the Walecka model \cite{Walecka:1974qa}, the vector channel $(\bar { \psi } \gamma ^ { \mu } \psi ) ^ { 2 }$ contribution is important at finite density. In the same way as the case for the finite density, we cannot neglect the contribution of the axial-vector channel $(\bar{\psi}i\gamma_5\gamma^{\mu}\psi)^2$ when we study the chiral imbalanced system. In previous NJL model analyses, people usually ignore the various channel contributions from the Fierz-transformed term or manually add the relevant terms. As shown below, the above mean field approximation approach for adding with finite chemical potential $\mu$ and chiral chemical potential is not self-consistent. In this paper, we employ a new self-consistent mean field approximation \cite{Wang:2019uwl,PhysRevD.100.043018} of NJL model to study the phase transition at nonzero chiral chemical potential and chemical potential. This model introduces a new free parameter $\alpha$ to reflect the proportion of the different channel contributions from the Fierz-transformed term and will lead to some new results. In Sec.$~\mathrm{\uppercase\expandafter{\romannumeral2}}~$, we will introduce the new self-consistent mean field approximation in the case of the chiral chemical potential and get the self-consistent gap equations. In Sec.$~\mathrm{\uppercase\expandafter{\romannumeral3}}~$,
we discuss the effects of chiral chemical potential on phase diagram and CEP with different $\alpha$. Finally, we will make a summary in Sec.$~\mathrm{\uppercase\expandafter{\romannumeral4}}~$.

\section{The New self-consistent mean field approximation of NJL model}
The standard two-flavour NJL Lagrangian with interaction term in the scalar and pseudoscalar-isovector channel is given by:
\begin{equation} \mathcal{L}_{NJL}=\bar{\psi} \left( i \slashed{\partial}-m_0 \right) \psi +G \left[\left(\bar{\psi}\psi\right)^2+\left(\bar{\psi} i \gamma_5\boldsymbol\tau \psi \right)^2\right] ,
\end{equation}
where $m_0$ is the current quark mass and $\psi=\begin{pmatrix}u&d\end{pmatrix}^T$ represents the quark fields.
We can perform the Fierz transformation on the four-Fermion interaction terms:
\begin{equation}\begin{split}
&\mathcal{L}_{IF}=\frac { G } { 8 N _ { c } } \left[ 2 ( \bar { \psi } \psi ) ^ { 2 } + 2 \left( \bar { \psi } i \gamma _ { 5 } \boldsymbol\tau  \psi \right) ^ { 2 } - 2 ( \bar { \psi } \boldsymbol\tau  \psi ) ^ { 2 } - 2 \left( \bar { \psi } i \gamma _ { 5 } \psi \right) ^ { 2 } \right.
\\&- 4 \left( \bar { \psi } \gamma ^ { \mu } \psi \right) ^ { 2 } -  4 \left( \bar { \psi } i \gamma ^ { \mu } \gamma _ { 5 } \psi \right) ^ { 2 }+ \left. \left( \bar { \psi } \sigma ^ { \mu \nu } \psi \right) ^ { 2 } - \left( \bar{ \psi } \sigma ^ { \mu \nu } \boldsymbol\tau  \psi \right) ^ { 2 } \right],
\end{split}\end{equation}
where $N_c$ is the number of colours. Then the Lagrangian becomes
\begin{equation}
\mathcal{L}_{F}=\bar{\psi}\left(i\slashed{\partial}-m_0\right)\psi+\mathcal{L}_{IF}.
\end{equation}
The Fierz transformation is a mathematical identity transformation, which relates the exchange and direct terms to each other. It is a very useful approach for us to understand clearly what this interaction is made of. And we can use Fierz transformation to derive a more general interaction term.

Although the original Lagrangian $\mathcal{L}_{NJL}$ and the Fierz-transformed Lagrangian $\mathcal{L}_{F}$ are identical, their contributions are no longer equivalent when we apply the mean field approximation. Especially in the case of an external field, the results yielded by $\mathcal{L}_{NJL}$ and $\mathcal{L}_{F}$ are quite different \cite{RevModPhys.64.649}. It means that it's important for us to know exactly the proportion of their contributions when we use the mean field approximation. The Ref. \cite{RevModPhys.64.649} employs the Lagrangian $\frac{1}{2}(\mathcal{L}_{NJL}+\mathcal{L}_F)$ , which is invariant under Fierz transformation. Therefore, the Hartree and Fock contributions of this Lagrangian are equal to each other. In fact, as the Refs. \cite{Wang:2019uwl,PhysRevD.100.043018} show, there is no physical requirement under the mean field approximation to ensure that Hartree and Fock contributions should be equal. 

Considering the $\mathcal{L}_{NJL}$ and $\mathcal{L}_F$ are mathematically equivalent, we employ the most general Lagrangian introduced by Refs. \cite{Wang:2019uwl,PhysRevD.100.043018} in this paper. The Lagrangian of the new self-consistent mean field approximation is
\begin{equation}\begin{aligned}
\mathcal{L}_R=(1-\alpha)\mathcal{L}_{NJL}+\alpha\mathcal{L}_F,
\end{aligned}\end{equation}
where the parameter $\alpha$ is an arbitrary $c$-number, which can be determined experimentally. As the $\mathcal{L}_{NJL}$ and $\mathcal{L}_F$ are equivalent, the Lagrangian doesn't change with $\alpha$.  But in the case of finite chemical potential, which can be regarded as the background vector field, we will get very different results from different $\alpha$ \cite{Wang:2019uwl,PhysRevD.100.043018}.

In order to study the effects of chiral imbalance between right-hand and left-hand quarks, we should add to the Lagrangian the following term \cite{Fukushima:2010fe,Fukushima:2008xe}:
\begin{equation}
\mu_5\bar{\psi}\gamma^0\gamma_5\psi.
\end{equation}
where $\mu_5$ is the chiral chemical potential coupling to the chiral density operator $\bar{\psi}\gamma^0\gamma_5\psi=\psi^{\dagger}_R\psi_R-\psi^{\dagger}_L\psi_L$. In this paper, we will discuss the effects of chiral chemical potential on phase transition at finite chemical potential. Therefore we have to employ the grand-canonical ensemble. The partition function is
\begin{equation}
Z=\operatorname{ Tr } e^{-\beta( H-\mu N ) }.
\end{equation}
In term of the Lagrangian, we can introduce the quark number density operator $\bar{\psi}\gamma^0\psi$ to it. Then the Lagrangian becomes
\begin{equation}
\mathcal{L}=(1-\alpha)\mathcal{L}_{NJL}+\alpha\mathcal{L}_F+\mu_5\bar{\psi}\gamma^0\gamma_5\psi+\mu\bar{\psi}\gamma^0\psi .
\end{equation}
In this paper, we only care about the scalar, vector and axial-vector channels contribution. Other terms have no effect on our studies at the level of mean field approximation. Applying the mean field approximation to this Lagrangian and dropping the irrelevant terms, we get the effective Lagrangian
\begin{equation}\begin{split}
\mathcal{L}_{\text{eff}}=&\bar{\psi}(i\slashed{\partial}-M+\mu^{\prime}\gamma^0+\mu^{\prime}_5\gamma^0\gamma_5)\psi-G\left(1-\alpha+\frac{\alpha}{4N_c}\right)\sigma^2
\\& +\frac{\alpha G}{2N_c}n^2-\frac{\alpha G}{2N_c}n_5^2 ,
\end{split}\end{equation}
where $M$ is often called "constituent quark mass":
\begin{equation}
M=m_0-2G \left(1-\alpha+\frac{\alpha}{4N_c}\right)\sigma,
\end{equation}
and
\begin{equation}\begin{split}
\mu^{\prime}&=\mu-\frac{\alpha G}{N_c}n,
\end{split}\end{equation}
\begin{equation}\begin{split}
\mu^{\prime}_5&=\mu_5+\frac{\alpha G}{N_c}n_5.
\end{split}\end{equation}
The quark condensation $\sigma=\langle\bar{\psi}\psi\rangle$, the quark number density $n=\langle\psi^{\dagger}\psi\rangle$ and the chiral number density $n_5=\langle\psi^{\dagger}\gamma_5\psi\rangle$ can be determined in a thermodynamically self-consistent way. The path integral representation of the partition function is
\begin{equation}
Z=\int_{\text{perodic}}\mathcal{D}\bar{\psi}\mathcal{D}\psi\exp \left( \int _ { 0 } ^ { \beta } d \tau \int d ^ { 3 } x \mathcal { L }_{\text{eff}} \right).
\end{equation}
And we can use the method introduced in the Ref. \cite{kapusta_gale_2006} to get the mean-field thermodynamic potential density
\begin{equation}\begin{split}
\Omega=&-\frac{T}{V} \ln Z
\\=&G\left(1-\alpha+\frac{\alpha}{4N_c}\right)\sigma^2-\frac{\alpha G}{2N_c}n^2+\frac{\alpha G}{2N_c}n_5^2+\Omega_{\text{M}},
\end{split}\end{equation}
where $\Omega_{\text{M}}$ is expressed as
\begin{equation}\begin{split}
\Omega_{\text{M}}=&-\frac{N_cN_f}{2\pi^2}\sum_{s=\pm 1}\int^{\Lambda}_0 p^2 \left\{\omega_s+T\ln [1+e^{-\beta(\omega_s+\mu^{\prime})}]\right.
\\&\left.+T\ln [1+e^{-\beta(\omega_s-\mu^{\prime})}]\right\} dp,
\end{split}\end{equation}
here $N_c=3$ and $N_f=2$ are respectively colour number and flavour number, the pole $\omega_s$ of the quark propagator is given by
\begin{equation}
\omega_s=\sqrt{M^2+(\mu^{\prime}_5-s|\mathbf{p}| )^2}
\end{equation}
and the index $s=\pm 1$ is the sign of the helicity.
In order to obtain the self-consistent equations of the thermodynamic equilibrium state, we have to find the minima of the thermodynamic potential density. Given the constraint conditions $\frac{\delta\Omega}{\delta\sigma}=0$, $\frac{\delta \Omega}{\delta n}=0$ and $\frac{\delta \Omega}{\delta n_5}=0$ of the minima, we can get the quark condensation:
\begin{equation}\begin{split}
\sigma=&\frac{\partial \Omega_{\text{M}}}{\partial M}
\\=&-M\frac{N_cN_f}{2\pi^2}\sum_{s=\pm 1}\int^{\Lambda}_0 \frac{p^2}{\omega_s}\left[1-f^+_s(p,\mu^{\prime},\mu_5^{\prime},T)\right.
\\&\left.-f^-_s(p,\mu^{\prime},\mu_5^{\prime},T)\right]dp,
\end{split}\end{equation}
the quark number density:
\begin{equation}\begin{split}
n=&-\frac{\partial \Omega_{\text{M}}}{\partial \mu^{\prime}}
\\=&\frac{N_cN_f}{2\pi^2}\sum_{s=\pm 1}\int^{\Lambda}_0 p^2\left[f^-_s(p,\mu^{\prime},\mu_5^{\prime},T)-f^+_s(p,\mu^{\prime},\mu_5^{\prime},T)\right] dp,
\end{split}\end{equation}
and the chiral number density:
\begin{equation}\begin{split}
n_5=&-\frac{\partial \Omega_{\text{M}}}{\partial \mu^{\prime}_5}
\\=&\frac{N_cN_f}{2\pi^2}\sum_{s=\pm 1}\int^{\Lambda}_0 p^2\frac{\mu^{\prime}_5-sp}{\omega_s}\left[1-f^-_s(p,\mu^{\prime},\mu_5^{\prime},T)\right.
\\&\left.-f^+_s(p,\mu^{\prime},\mu_5^{\prime},T)\right] dp,
\end{split}\end{equation}
where $f^{\pm}_s$ is the Fermi-Dirac distribution represented as
\begin{equation}
f^{\pm}_s(p,\mu^{\prime},\mu_5^{\prime},T)=\frac{1}{1+e^{\beta(\omega_s\pm\mu^{\prime})}}.
\end{equation}
Finally, plugging the Eqs. (17-19) into the Eqs. (10-12), we will obtain the self-consistent gap equations in the case of finite $\mu$ and $\mu_5$.

In order to fit the experimental results (the pion mass $m_\pi=138\ \text{MeV}$, decay constant $f_\pi=93\ \text{MeV}$ and quark condensate per flavour $\langle\bar{\psi}\psi\rangle=-(250\ \text{MeV})^3$), which are obtained at $T=~0$, $\mu=0$ and $\mu_5=0$, we must redefine the coupling constant $G$ of this new self-consistent method as
\begin{equation}
G=\frac{1+\frac{1}{4N_c}}{1-\alpha+\frac{\alpha}{4N_c}}g,
\end{equation}
where $g=5.074\times10^{-6}\,\text{MeV}^{-2}$ is the coupling constant of the conventional mean field approximation of NJL model \cite{Hatsuda:1994pi}. The remaining parameters are the cut-off $\Lambda=631\ \text{MeV}$ and the current quark mass $m_0=5.5\ \text{MeV}$ \cite{Hatsuda:1994pi}. As indicated in Refs. \cite{Wang:2019uwl,PhysRevD.100.043018}, $\alpha$ cannot be given in advance by the mean field theory, it must be determined by a finite density experiment. For example, it can be determined by astronomical observation data on the latest neutron star merger \cite{PhysRevD.100.043018}. In this paper, we consider $\alpha$ as a free parameter to see the effect of chiral chemical potential $\mu_5$ on QCD phase structure under different $\alpha$ values.

\section{Numerical Calculations And Analysis}
First, we consider the condition of $\alpha=0.5$. We solve the gap equations Eqs. (10-12) numerically and obtain the phase diagram Fig. 1 on the $\mu-T$ plane with different $\mu_5$.\footnote{Given that we use the 3-momentum hard cut-off regularization scheme in this paper, T, $\mu$ and $\mu_5$ are limited by the cut-off $\Lambda$. It is sensible to keep the values of T, $\mu$, $\mu_5$ lower than 500 MeV. }  As we can see, the chiral chemical potential $\mu_5$ could reduce the critical chemical potential $\mu_c$ of phase transition at a low temperature.  When the chiral chemical potential is zero, the critical end point (CEP) locates at $(\mu,T)^{CEP}=(340.1,44) \ \text{MeV}$ which is quit different with the result $(160,165) \ \text{MeV}$ of Ref. \cite{Ruggieri:2011xc}. The change of CEP with $\mu_5$ is similar to the Ref. \cite{Lu:2016uwy}, in which the vector interaction is not involved. The temperature of CEP increases with the chiral chemical potential up to a maximum of $T^{CEP}_{max}=96\ \text{MeV}$, and afterwards it decreases. The chemical potential of CEP always decreases with increasing $\mu_5$. When the temperature is higher than $T^{CEP}_{max}$, the phase transition is crossover with all $\mu_5$.

\begin{figure}[b]
\centering
	\includegraphics[width=1\linewidth]{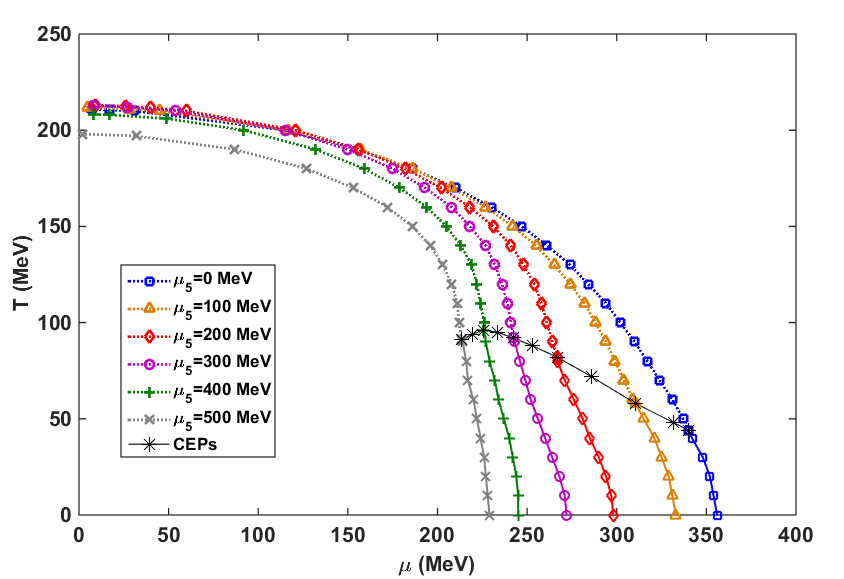}
	\caption{ The phase diagram of $\alpha=0.5$. The solid lines and the dotted lines respectively represent the first-order phase transition and crossover at the corresponding $\mu_5$. The stars denote the CEPs. The difference in $\mu_5$ between each star is 50 MeV.}
\end{figure}

The phase diagram of $\alpha=0.8$ at different chiral chemical potential is shown in Fig. 2. When $\mu_5=0$, the phase transition is crossover at zero temperature and there is no CEP on the phase diagram. As stated in the Ref. \cite{Wang:2019uwl}, when $\alpha$ is greater than the critical value 0.71, there is no CEP with $\mu_5=0$. As the chiral chemical potential increases, there is not CEP until $\mu_5=54\ \text{MeV}$. When $\mu_5=54\ \text{MeV}$, at zero temperature, the phase transition is no longer crossover but a first-order phase transition and the CEP appears at $(\mu,T)^{CEP}\approx(359.2\ ,\ 0)\ \text{MeV}$. The temperature of CEP increases at first. After reaching the maximum $T^{CEP}_{max}\approx 69\ \text{MeV}$, it decreases. 
   	
\begin{figure}[t]
\centering
	\includegraphics[width=1\linewidth]{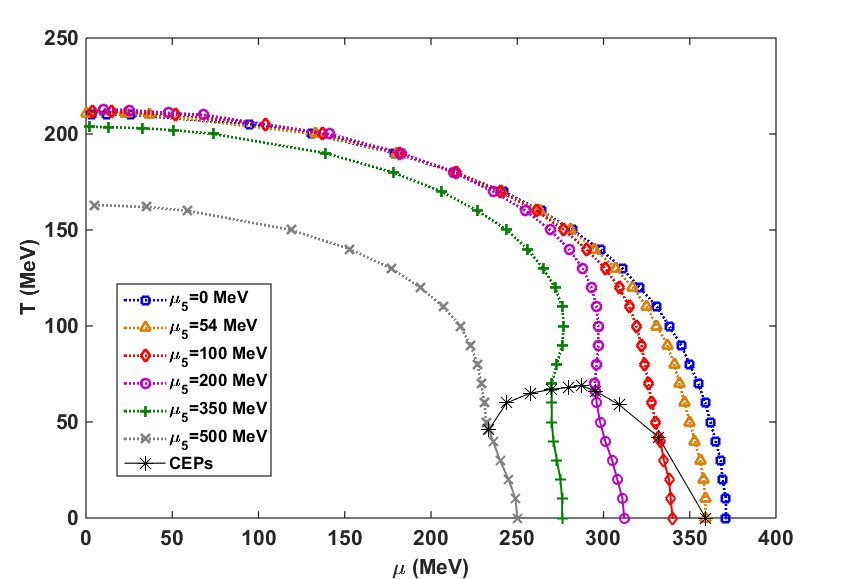}
	\caption{ The phase diagram of $\alpha=0.8$. The solid lines and the dotted lines respectively represent the first-order phase transition and crossover at the corresponding $\mu_5$. The stars denote the CEPs. The difference in $\mu_5$ between each star is $50\ \text{MeV}$.}
\end{figure}	
\begin{figure}[b]
\centering
	\includegraphics[width=1\linewidth]{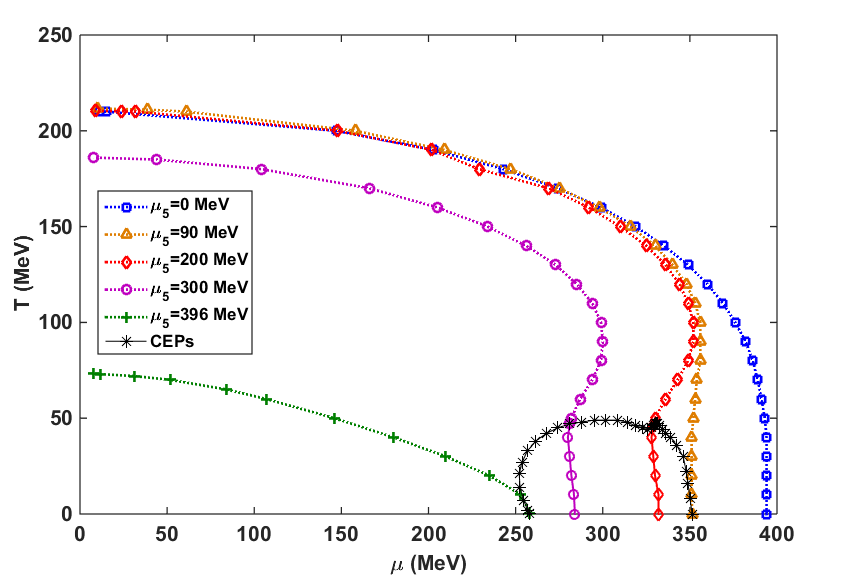}
	\caption{ The phase diagram of $\alpha=0.9$. The stars denote the projection of CEPs onto $\mu-T$ plane. The solid lines and the dotted lines respectively represent the first-order phase transition and crossover at the corresponding $\mu_5$. }
\centering
\end{figure}

Similar results appear for other values of $\alpha$ greater than 0.71. We can see from the phase diagram Fig. 3 of $\alpha=0.9$ that the phase diagram does not have CEP with $\mu_5=0$. When the chiral chemical potential increases to the critical value $90\ \text{MeV}$, the CEP appears on the $\mu$ axis at $\mu=351.3\ \text{MeV}$. We display the critical temperature $T_c$ of phase transition at $\mu=0$ as a function of $\mu_5$ for all the presented values of $\alpha$ in Fig. 4.  As we can see, at a high temperature, as the chiral chemical potential rises, the critical temperature $T_c$ of phase transition increases slowly at the beginning, and then decreases quickly. The projection of CEPs onto $\mu_5-\mu$ plane is plotted in Fig. 5. When $\alpha=0.9$, as the chiral chemical potential grows, the chemical potential of CEP decreases when $\mu_5<360\ \text{MeV}$ but increases when $\mu_5>360\ \text{MeV}$. The minimum of the chemical potential of CEP is $\mu^{CEP}_{min}=252.1\ \text{MeV}$. It means that the phase transition is totally crossover and there is no CEP on $\mu_5-T$ plane when $\mu \textless 252.1\ \text{MeV}$. This result contradicts the calculation of PNJL model \cite{Ruggieri:2011xc}, which predicts that the CEP can consecutively move to the $\text{CEP}_5$ (the CEP in $\mu_5-T$ plane at $\mu=0$). The projection of CEPs onto the $\mu_5-T$ plane is plotted in Fig. 6. It shows us that, when $\alpha=0.9$, the CEP appears at $\mu_5\approx 90\ \text{MeV}$ and vanishes at $\mu_5\approx 396\ \text{MeV}$. That is to say the phase transition is crossover when $\mu_5< 90\ \text{MeV}$ or $\mu_5>396\ \text{MeV}$. There are two peaks of temperature at $\mu_5\approx160\ \text{MeV}$ and $\mu_5\approx270\ \text{MeV}$. And the maximum temperature of CEP is $T^{CEP}_{max}\approx 49\ \text{MeV}$ at $\mu_5\approx270\ \text{MeV}$. This result indicate that, if the temperature is higher than $T^{CEP}_{max}$, the phase transition must be crossover.

\begin{figure}[t]
\centering
     \includegraphics[width=1\linewidth]{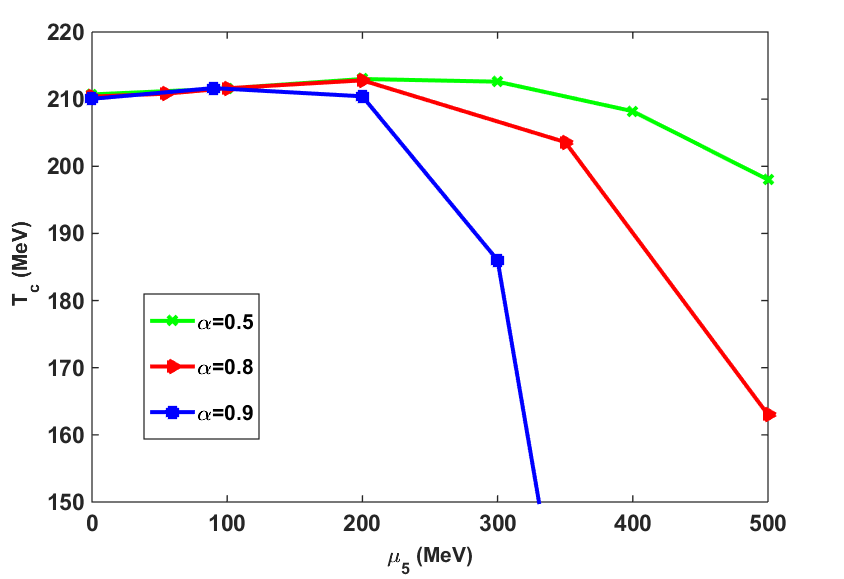}
	\caption{ The critical temperature $T_c$ of phase transition changes with $\mu_5$ at $\mu=0$ for different $\alpha$.  }
\end{figure}		
\begin{figure}[b]
	\includegraphics[width=1\linewidth]{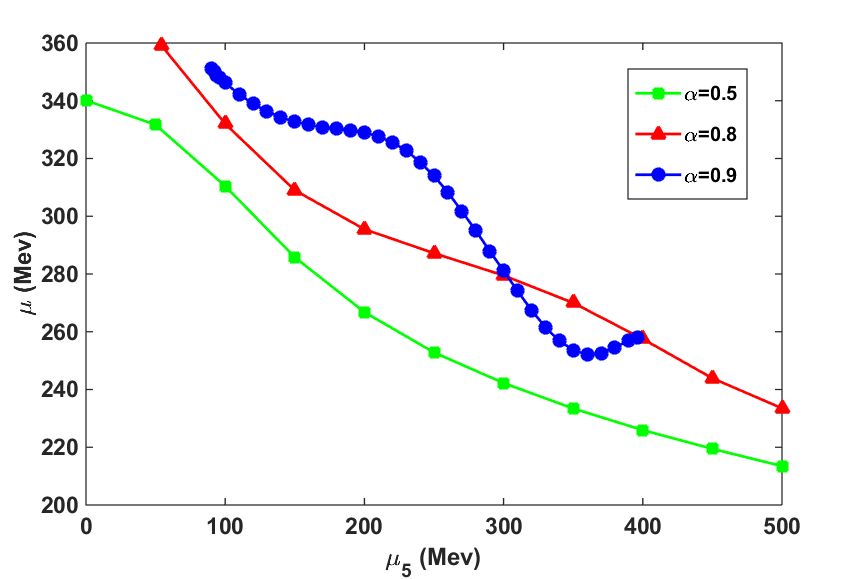}
	\caption{ The projection of CEPs onto $\mu_5-\mu$ plane for different $\alpha$.}
\end{figure}	

With different $\alpha$, phase diagram changes with $\mu_5$ in a similar way. 
As shown in Fig. 4, at a high temperature, the chiral chemical potential will raise the critical temperature $T_c$ of phase transition at first, but when $\mu_5>200\ \text{MeV}$, $T_c$ will quickly decrease. Under the condition of low temperature and high density, the chiral chemical potential will lower the critical chemical potential $\mu_c$ of phase transition. As we can see in Fig. 6, the temperature of CEP increases with $\mu_5$ growing at first, undergoes  a long plateau around the maximum and then decreases. When $\alpha>0.71$, there is no CEP with $\mu_5=0$. As the chiral chemical potential increases, the CEP reappears. What needs to be pointed out here is that as far as we know, this is a completely new result and we have never seen a similar report before. 

\begin{figure}[t]
	\centering
	\includegraphics[width=1\linewidth]{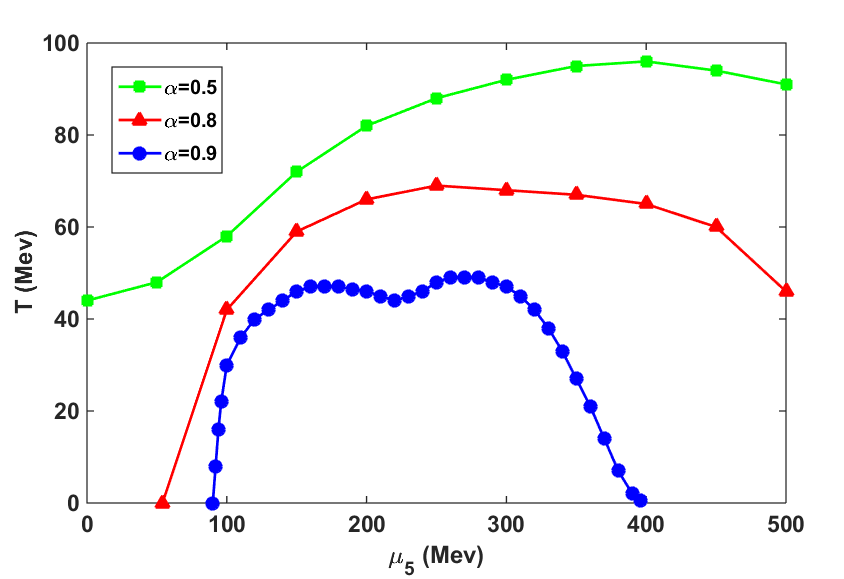}
	\caption{ The projection of CEPs onto $\mu_5-T$ plane for different $\alpha$.}
\end{figure}	

\section{Conclusion}
In this paper, we employed the new self-consistent mean field approximation to study the QCD phase diagram in the case of chiral imbalance. Our results show that, at high temperature, the critical temperature $T_c$ of phase transition rises slowly with increasing the chiral chemical potential, but when $\mu_5>200\ \text{MeV}$,  $T_c$ will decrease rapidly. This result is consistent with the Ref. \cite{PhysRevD.94.014026}, which uses three different regularization schemes to study the effects of chiral chemical potential. In relativistic heavy-ion collisions, the chiral chemical potential is estimated to be about 10 \textasciitilde 100 MeV \cite{Hirono:2014oda,Jiang:2016wve,Lin:2018nxj}. As shown in Fig. 4, within this range, the chiral chemical potential will raise the critical temperature $T_c$ slowly. At the low temperature, as the chiral chemical potential increases, the critical chemical potential $\mu_c$ will diminish. 
The location of CEP is studied in this paper and we found that the temperature of CEP will increase at first, reaches the maximum $T^{CEP}_{max}$ then decrease with increasing of $\mu_5$. When the temperature is higher than $T^{CEP}_{max}$, the phase transition becomes crossover. If the temperature is lower than $T^{CEP}_{max}$, the crossover will become a first-order phase transition.
The previous works show that when $\alpha>0.71$ \cite{Wang:2019uwl,PhysRevD.100.043018}, there is no CEP on the QCD phase diagram. But as we have shown in this paper that, in the chiral imbalanced system, the CEP will appear again with increasing chiral chemical potential $\mu_5$. As shown in Ref. \cite{Lin:2018nxj}, the chiral chemical potential shows a positive correlation with centrality. It is possible to adjust the centrality to change the chiral chemical potential. This means that we may possibly regulate the chiral imbalance in the rela\!\,tivistic heavy-ion collision experiment to better observe the possible CEP signal. 

\begin{acknowledgments}
This work is supported in part by the National Natural Science Foundation of China (under Grants No. 11475085, No. 11535005, and No. 11690030) and by Nation Major State Basic Research and Development of China (2016YFE0129300). X. Luo is supported in part by the National Natural Science Foundation of China under Grants (No. 11890711, 11575069, 11828501 and 11861131009), Fundamental Research Funds for the Central Universities No. CCNU19QN054. 
\end{acknowledgments}

\bibliographystyle{apsrev4-1}

\bibliography{ref}

\end{document}